\documentclass{article}

\usepackage[english]{babel}

\usepackage[letterpaper,top=2cm,bottom=2cm,left=3cm,right=3cm,marginparwidth=1.75cm]{geometry}

\usepackage{amsmath}
\usepackage{graphicx}
\usepackage[colorlinks=true, allcolors=blue]{hyperref}

\title{Modeling Complex Multiphysics Systems with Discrete Element Method Enriched with the Kernel-Independent Fast Multipole Method}

\author{Igor A. Ostanin}

\begin{document}
\maketitle

\begin{abstract}
The paper describes the coupling of the MercuryDPM discrete element method (DEM) code and the implementation of the kernel-independent fast multipole method (KIFMM). The combined simulation framework allows addressing the large class of multiscale problems, including both the mechanical interactions of particulates at the fine scale and the long-range interactions of various natures at the coarse scale. Among these are electrostatic interactions in powders, clays, and particulates, magnetic interactions in ferromagnetic granulates, and gravitational interactions in asteroid clouds. 
The formalism of rigid clumps is successfully combined with KIFMM, enabling addressing problems involving complex long-large interactions between non-spherical particles with arbitrary charge distributions.
The capabilities of our technique are demonstrated in several application examples.      
\end{abstract}

\section{Introduction}

The discrete element method (DEM) \cite{Cundall1979} was originally proposed by P. Cundall as a tool to model the mechanical properties of granular solids. It is based on a rather powerful and flexible modeling paradigm, \textit{i. e.}, the representation of a mechanical system of interest as the set of interacting rigid bodies. Such a model can be used to study complex dynamic phenomena or, alternatively, to find the equilibria of complex mechanical systems. DEM was rapidly developing over four decades, expanding to multiple new areas of application. DEM-based approaches proved to be efficient in addressing the challenges of wide particle-size distributions \cite{Ogarko2012}, complex nonspherical shapes of the particles \cite{Lu2015} and their deformability \cite{Ghods2022}, incorporation of complex contact mechanics \cite{BroduDijksmanBehringer2015}, and the predictive modeling of soft matter bound by van der Waals interactions \cite{Ostanin2013, Grebenko2022}. The existing DEM projects (\textit{e. g.} \cite{Yade_2021,Govender_2016,weinhart2020fast}) demonstrate the maturity and efficiency of their architectures, high performance, cross-compatibility, and efficient parallelization on different high-performance computing architectures. 

However, one serious limitation of DEM is its inability to tackle the problems involving the long-range interactions. The core of any DEM code is the employed procedure for the detection of interactions between particles. Different DEM codes employ different procedures: the codes oriented towards dense particle assemblies \cite{Yade_2021} usually employ variants of the sweep-and-prune algorithm \cite{Cohen1995}, while the codes oriented towards modeling particle flows \cite{weinhart2020fast} utilize variations of the linked cell algorithm \cite{Ogarko2012}, etc. However, in any case, those algorithms are designed for efficient detection of short-range (mechanical) interactions.
Analysis of the long-range interactions is essentially a different problem that requires different approaches to approximation of interactions, domain partitioning, parallelization, etc. As a result, there are currently no codes that efficiently combine DEM with long-range interactions.

If particles interact via the long-range forces of some nature, and this interaction can not be neglected, one needs to account for all possible pair interactions, which results in $\mathcal{O}(N^2)$ complexity of forces computation. However, in practice, it is usually sufficient to evaluate the forces with some prescribed precision. This problem admits approaches that are asymptotically faster, providing either $\mathcal{O}(N \log N)$ or even $\mathcal{O}(N)$ complexity. Among those are Barnes–Hut tree algorithm \cite{Barnes1986}, the Fast Multipole Method (FMM) \cite{Greengard1987}, and subsequent hierarchical matrix frameworks including $\mathcal{H}$-, $\mathcal{H}^2$-, and HSS-matrices \cite{Hackbusch1999, Hackbusch2000, Chandrasekaran2006} providing scalable approximations for N-body problems.

On the other hand, a similar family of fast summation methods can be used to efficiently accelerate boundary integral equation (BIE) formulations by factorizing the linear systems that arise after discretization (e.g., \cite{Abtin2011, Ostanin2017}). This approach enables a reduced–dimension discretization of the governing PDE (surface rather than volume), achieves low asymptotic complexity per iteration ($\mathcal{O}(N)$), and yields well-conditioned linear systems, allowing accurate solutions in only a few GMRES iterations (see, e.g., \cite{KeavenyShelley2011, ChandlerWildeSpence2024}). When coupled with DEM --— i.e., when particle dynamics are solved concurrently with the BIE on particle surfaces —-- the range of feasible applications expands significantly. The resulting DEM–fast-summation framework can be applied to diverse scenarios, including viscous, elastic, and viscoelastic continua surrounding particles, as well as interactions induced by dielectric polarization, magnetization, and related effects. Overall, this enrichment of DEM opens a path toward solving longstanding problems in the powders-and-grains community that have remained untouched for decades.

Important requirements for a fast summation framework suitable for such DEM extension include: i) support of different kernels, preferably, without the need for manual computation of multipole expansions of those; ii) scalable parallel implementation; iii) low compute cost of a single field evaluation (or matrix-vector product in case of BIE solution). We have chosen a Kernel-Independent Fast Multipole Method \cite{pvfmm} as an instrument that meets all these requirements. 

In the current paper, we do not explore the use of the coupled framework in the context of solving coupled DEM-BIE problems; the focus of this work is on accelerating the DEM simulations with long-range interactions. A special attention is paid to simulation of aspherical particles modeled as rigid clumps \cite{Thornton2023, Ostanin2024RigidClumps} in presence of long-range forces simulated using KIFMM. In the remainder of the paper we will discuss the technical features of a coupled code and illustrate its capabilities with a number of examples.

\section{Methods}

\subsection{\textit{MercuryDPM} particle dynamics code}

\textit{MercuryDPM} \cite{weinhart2020fast, MDPMBitbucket} is an open-source implementation of DEM used to simulate granular materials —-- systems composed of individual particles found in a wide range of natural and industrial contexts. Typical examples include sand, coffee beans, iron ore pellets, pharmaceutical tablets, catalysts, etc. Understanding the behavior of these materials is essential for various industries, such as pharmaceuticals, mining, food processing, and manufacturing.

Development of the code began in 2009 at the University of Twente, and it has since evolved into a comprehensive framework supported by a large open-source community of academic and industrial users. \textit{MercuryDPM} is a modular, object-oriented C++ software package built and tested using \texttt{cmake/ctest}.

The software includes three key features that make it suitable for modeling complex industrial and natural systems:

\begin{itemize}
    \item a flexible architecture allowing for intricate wall geometries and boundary conditions;

    \item an integrated analysis toolkit capable of efficiently processing large simulation datasets; and

    \item an advanced contact detection algorithm that ensures high computational efficiency, particularly for systems with strongly polydisperse particles \cite{Ogarko2012, Mercury2020}.
\end{itemize}

By default, \textit{MercuryDPM} represents particles as spheres (discrete elements) defined by their mass, radius, position, velocity, and angular velocity. It also provides support for superquadric particles \cite{Mercury2020} and rigid clumps \cite{Ostanin2024RigidClumps} that will be briefly reviewed below. A typical simulation scenario involves the computation of damped dynamics of multiple particles, employing the Velocity Verlet algorithm to update particle positions and the forward Euler scheme for rotational motion. A wide library of (short-ranged) contact models is available to describe the physical laws governing normal and tangential forces during particle interactions --- e.g. Hertzian contacts, liquid bridges, van der Waals forces etc.

\subsection{Kernel-independent fast multipole method}

We utilize KIFMM to enrich a DEM particle simulation with long range forces of various natures. KIFMM was first proposed in 2004 \cite{kifmm}; in our work we utilize the library pvfmm \cite{pvfmm} --- an active project that builds on the original KIFMM implementation by L. Ying and co-authors.

For completeness, we outline here main ideas of fast multipole methods. In order to reach linear (with respect to the number of particles $N$) complexity and elapsed time of computing of long-range pair interactions, fast multipole methods perform the following steps\cite{fmm_tut}:

\begin{itemize}
	\item  Generation of an octree partitioning of the domain into boxes, so that every leaf box contains not more than $M$ particles ($M<<N$).
    
	\item  Fine-to-coarse tree traversal to compute compact representations of the far-field potential of a box (\emph{multipole expansions} for analytic FMM); these are computed hierarchically, by combining expansions of descendant boxes to the expansion for the parent box, using linear M2M (``multipole-to-multipole'' translation operators. Multipole expansions are used to approximate the values of the integral over all points contained in a box, with the evaluation point far enough away from that box;
          
	\item  a coarse-to-fine pass, that computes \emph{local expansions} for each box, that approximate the values of the integral over all points far enough away from the box --- inside the box. These are obtained at descendant boxes by combining the parent's local expansion (using an L2L, ``local-to-local'' operator) with multipole expansions of boxes that are not in the far zone of the parent, but are in the far zones of the descendants.  Multipole expansions are converted to local using M2L operators.   
	\item  At the finest level of the tree, the complete integrals are computed by adding the far-field and the contributions of points in the near zone.           
\end{itemize}

The distinguishing feature of KIFMM, compared to the original analytic FMM method is that it does not require explicit multipole and local series expansions of underlying kernels, and analytically derived M2M, M2L, L2L operators for each kernel.
Instead, it represents the far-field (multipole) and local approximations of the integrals with a density $\phi$  defined at samples $x_i$ of an equivalent surface, so that the approximation at a point $y$  has the form $\sum_i \phi_i K(x_i,y)$, where $K(x,y)$ is the kernel of interest.  

The M2M, L2L and M2L operators needed in the algorithm, in the case of KIFMM are represented by matrices mapping density values on different equivalent surfaces to each other, and are computed automatically for each needed kernel. 

Just like the original FMM, kernel-independent FMM performs the summation of the field of $N$ sources/targets with $O(N)$ operations (in case of summation of pairwse interactions ``sources'' and ``targets'' are identical).
In our work we use a recent parallel implementation of KIFMM --- PVFMM \cite{kifmm,pvfmm}, which implements rapid evaluation of sums of the following 
\begin{equation} \label{fmm_summation}
t_{i}( \mathbf{x}_{i})=K_{ij}(\mathbf{x}_{i}, \mathbf{y}_{j})s_{j}(\mathbf{y}_{j})
\end{equation}     
$t_{i}$ is the vector of target values being computed (values of at points of interest $\mathbf{x}_{i}$);  $s_{j}$ is the vector of known source values at points $\mathbf{y}_{j}$ (solution values on the surface). Source and target values may be scalars (e.g., charges and potentials), vectors (traction forces and target displacements) or higher-rank tensors. Kernel function $K_{ij} (\mathbf{x}_{i}, \mathbf{y}_{j},\mathbf{n}_{j})$  depends on both source and target coordinates; it can also be a scalar, vector or tensor; for some kernels it may depend on normals at the source and the target points. In this paper we will showcase the summation of scalar Laplacian potential and its vector gradient (force).

\subsection{Rigid clumps}

One of the essential features of MercuryDPM, useful in the context of particles with long-range interactions are rigid clumps \cite{Thornton2023, Ostanin2024RigidClumps}. By \textit{rigid clump} (or just \textit{clump}) we will imply an aggregate of $N$ rigid spherical particles of a given density, that are rigidly linked to each other at a given relative translational and rotational positions (Fig. 1). The constituent particles of a clump will be referred to as \textit{pebbles}. The clump is a \textit{rigid body} possessing $6$ degrees of freedom. Rigid clump of pebbles has its collective mass and tensor of inertia (TOI) and treated as a single particle by the time integration algorithm  --- and as a collection of particles by the contact detection algorithm. 

Rigid clumps provide a very useful tool to create particles with complex charge distribution. By the particle's charge we will imply any kind of density of source value in sum \ref{fmm_summation}. Every pebble can be assigned a certain density; alternatively, only few ``basis'' pebbles can be assigned a density approximating a complex charge distribution --- fast techniques for finding the basis densities are described in \cite{MikhalevOseledets2016}.

\subsection{Coupling \textit{MercuryDPM} Drivers with the \textit{KIFMM} Library}

Because both \textit{KIFMM} and \textit{MercuryDPM} employ the CMake build system, their integration is relatively straightforward. \textit{MercuryDPM} provides a dedicated compilation option enabling linkage with \textit{KIFMM}, which is incorporated as a static library within the \textit{MercuryDPM} drivers.

To accommodate long-range particle interactions, the abstract particle class is augmented with an additional attribute --- charge --- represented by a single real-valued parameter. The force computation pipeline is correspondingly extended by an additional stage responsible for accumulating long-range interaction forces.

For each driver, the problem class is derived from the standard \texttt{Mercury3D} class. Within the \texttt{setupInitialConditions()} method, the system initializes the \textit{KIFMM} component, including the construction of the initial tree structure and the precomputation of translation operators corresponding to the selected kernels. The long-range force (and the corresponding contribution to the potential energy) evaluation is executed in the \texttt{actionsAfterTimeStep()} method, where KIFMM computes the contributions of long-range interactions for all particles. For clumped particles, charges are distributed across pebble constituents, and the resulting forces are first applied at the pebble level. Subsequently, all forces and moments acting on individual pebbles are aggregated to obtain the total force and torque on each clump.

In worth noting that the maximum stiffness of mechanical collisions $k_{DEM}$ and long-range interactions $k_{LR}$ typically differ by several orders of magnitude. As the Rayleigh time step scales with the square root of stiffness, it is reasonable to re-compute the long-range additions not less than once after $S$ timesteps, where

\begin{equation} \label{fmm_stride}
S \approx \sqrt{\frac{k_{DEM}}{k_{LR}}}
\end{equation}     

We will hereafter call this parameter a long-range force computation stride. 

Periodic boundary conditions can be applied in both the MercuryDPM contact-detection algorithm and in KIFMM to model an effectively infinite, tiled domain. However, when summing long-range interaction kernels over periodic images the lattice sums only converge under certain regularity (neutrality) conditions: for the Laplace/Coulomb kernel \cite{deLeeuw1980} the condition is that the total charge in a single periodic cell vanishes (otherwise the potential or energy diverges, or an artificial uniform neutralizing background is implicitly introduced). More generally, different kernels impose different moment-constraints (for example, Stokes flows require zero net force and torque for a well-posed periodic sum).

The KIFMM computation, which is typically the most time-consuming part of the timestep, is executed in MPI-parallel mode, whereas the \textit{MercuryDPM} driver employs shared-memory parallelism via OpenMP. This hybrid parallel architecture ensures adequate performance for simulations extending over $10^6$ timesteps and involving up to $10^6$ particles.

The current implementation is sufficiently flexible to support multiple extensions, including the incorporation of boundary-integral-equation (BIE) solvers for particle surfaces or a fully concurrent MPI-parallelized coupling of DEM and KIFMM. 

Overall, combination of DEM, rigid clumps and long range interactions results in a powerful modeling framework, whose performance is highlighted below with a few interesting examples.

\section{Results and Discussion}

In this section, we present several illustrative examples of MercuryDPM–KIFMM simulations. These examples demonstrate the capabilities of our modeling approach and provide a concise overview of the classes of problems that can be addressed with our simulation framework. As the examples cover a large span of length and time scales, physics areas and qualitative characteristics of the processes under consideration, we do not offer a unified system of units for every example --- instead, we use the following simple normalizations. For every simulation the time is given in the units of simulation duration ($\hat{t} = T/T_{max}$), while the energies are expressed in units of reference energy state $\hat{U} = U / U_{ref}$, where $U_{ref}$ is discussed separately for every example. 

All simulations described in this chapter were performed on an HP Z2 workstation equipped with 32 GB of RAM and a 16-core Intel i9 CPU. KIFMM used hybrid intra-node OMP/MPI parallelization, whereas MercuryDPM relied solely on intra-node OMP parallelization. The reported simulation times are approximate and include data-processing and I/O overheads.

The driver codes containing the complete set of simulation parameters are available in the project repository (see below).

\subsection{Identical Charged Particles in a Box}

This example illustrates our ability to simulate the effects of electric charge in granular systems, which are notoriously hard to model, but have extreme practical importance, in particular, for questions of powder's processing safety \cite{Nifuku2003} and effective mechanical properties \cite{Karner2011}. We consider $10^4$ identical, uniformly charged particles of radius $r = 0.0025$, interacting through a repulsive electrostatic force

\begin{equation} \label{electorstatic}
\mathbf{F}_{ij} = \mathbf{r}\,\frac{1}{4\pi}\,\frac{q_i q_j}{r^3}
\end{equation}     

In addition to the long-range interaction, a short-range contact model is applied at every particle–particle and particle–wall interaction. This model uses the linear viscoelastic frictional formulation implemented in MercuryDPM~\cite{Mercury2020}. Small viscous dissipation, as well as sliding and rolling friction, ensures that the system relaxes to equilibrium.

The initial particle centers are uniformly distributed in the cube $(r,\,1-r)^3$. The faces of the unit cube act as elastic walls, preventing particles from leaving the computational domain. The model parameters guarantee stable numerical integration with a long-range force–evaluation stride of $S = 10$.

After a brief transient, all particles migrate to the walls of the box, with higher concentrations near the edges and corners as expected. Figure 1(A) shows snapshots at $\hat{t} = 0$, $\hat{t} = 0.05$, and $\hat{t} = 1$. Figure 1(B) displays the evolution of the long-range potential energy (the reference state $U_{ref} = U_{LR}(0)$). The dynamics exhibit a fast initial stage --- particle deposition onto the walls --- followed by a slower rearrangement along the walls. Figure 1(C) shows the final equilibrium configuration of particles on one of the walls. Notably, the resulting packing is not crystalline for the parameters considered, likely due to strong edge effects.

A simulation of this scale took 2 hours on a CPU node. The precise simulation parameters can be found in \texttt{Drivers/LongRange/Charges/}.

\begin{figure} 
    \centering
    \includegraphics[width=\textwidth]{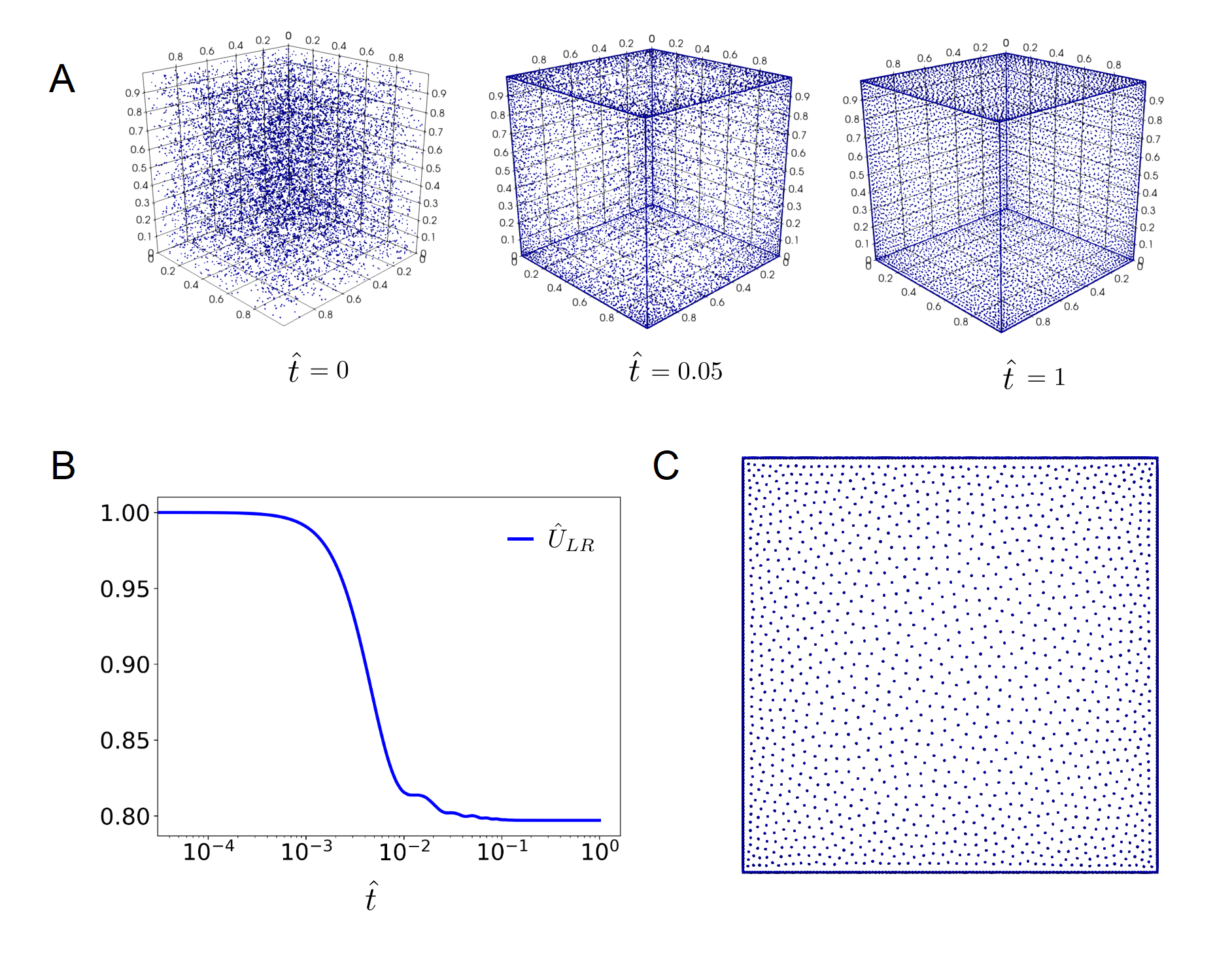}
    \caption{Identical charged particles in the box. (A) Snapshots of the system evolution at the beginning of the simulation ($\hat{t} = 0$), after $2 \times 10^4$ timesteps ($\hat{t} = 0.05$), and after $10^5$ timesteps ($\hat{t} = 1$). (B) Time evolution of the potential energy of long-range interactions $U_{LR}$ during the simulation. (C) Final stable configuration of particles at one of the sides of the box. }
    \label{fig:2}
\end{figure}

\subsection{Aggregation of interacting dipoles}

The second example demonstrates the capabilities of our simulation framework to model complex electrochemical and electrotribological processes involving dipole interactions between particles of powders and other soft materials, similar to \cite{Satoh2016, Domingues2018}. Consider a system of $2 \times 10^4$ rigid clumps, each represented by three  spherical pebbles of the radius $r = 0.005$ equispaced by $l = 0.01$ with charges $-1$, $0$, and $1$, respectively. The mechanical model of interactions between pebbles is identical to the one in the first example. The dipoles are confined in a unit box $(l+r,1-(l+r))^3$ with elastic walls and are subjected to an external homogeneous gravitational field. A sufficient background drag is added to model the effect of the surrounding fluid medium. 

Our simulation features aggregation of the dipoles into several nearly rigid dendritic aggregates and their slow deposition onto the bottom of the box. Similarly to adhesion of rod-like structures due to short-ranged vdW interactions \cite{Ostanin2018}, we can see the monotonous decrease of adhesion energy reaching plateau, associated with formation of stable network of aggregates. However, the individual particles tend to form much larger aggregates, as the interaction radius is much larger than the size of the particle. Figure 1(A) shows several snapshots highlighting the aggregation process. Figure~2(B) presents the corresponding evolution of several terms in the total energy of the system ($U_{ref} = |U_{LR}(1)|$). 

One can see that the initial overlaps between clumps create an initial peak in elastic energy, which immediately drops to nearly zero; thereafter, the elastic energy of collisions is negligible compared with other potential energy terms. The background drag leads to fast quenching of the initial kinetic energy. The gravitational energy of the system decreases as the aggregates settle onto the bottom surface of the elastic box. The cumulative energy of dipole–dipole interactions, $U_{LR}$, features monotonous degrease till certain equilibrium level. The aggregation process ends as the potential energy reaches the local minimum associated with a particular topology of dendrites.

The showcased simulation took 14 hours on a CPU node. The simulation parameters can be found in \texttt{Drivers/LongRange/DipolesAggregation/}.

\begin{figure} 
    \centering
    \includegraphics[width=\textwidth]{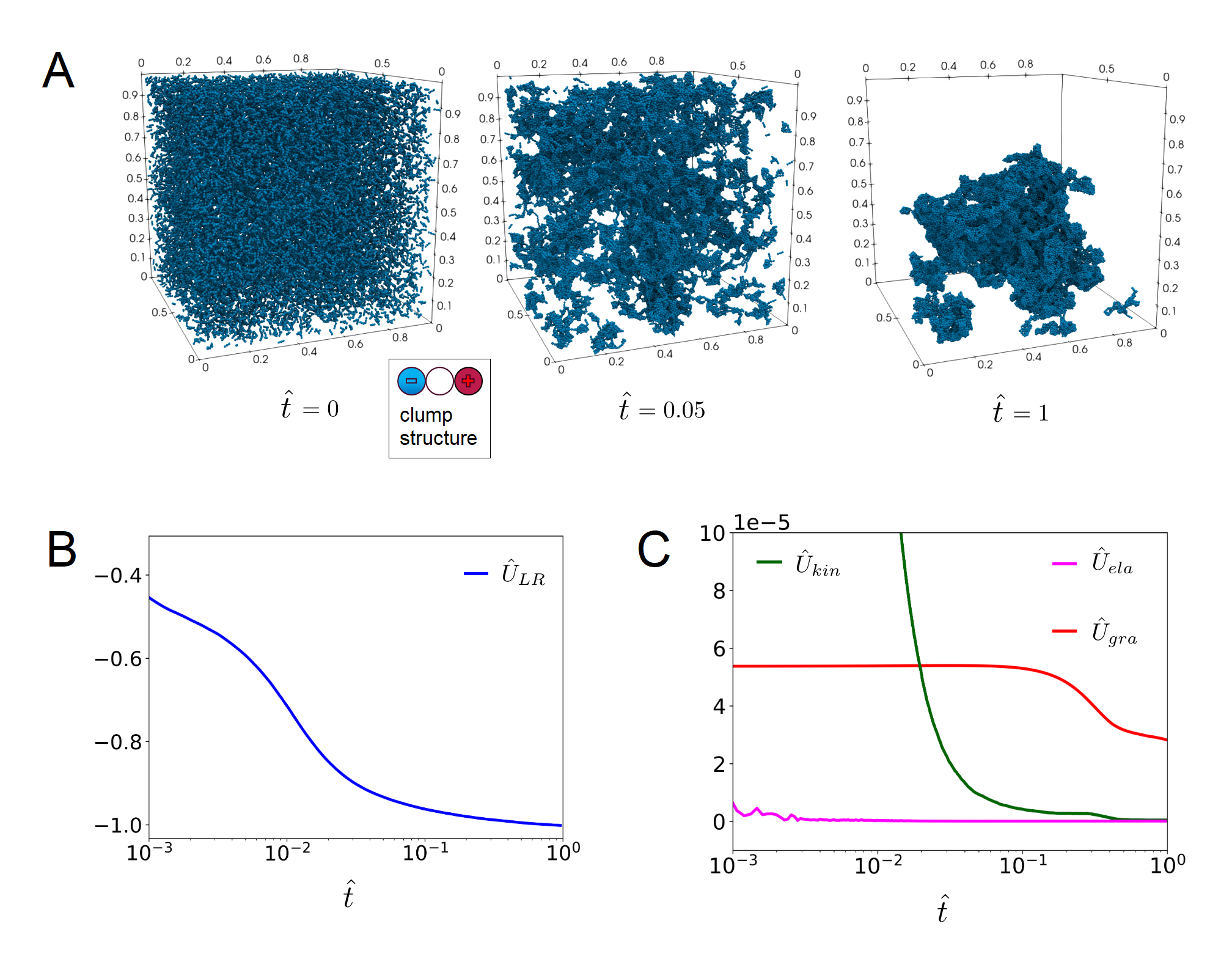}
    \caption{Aggregation of interacting dipoles. (A) Snapshots of the system at $\hat{t} = 0$, $\hat{t} = 0.05$, $\hat{t} = 1$. (B) Time evolution of the potential energy of long-range interactions $U_{LR}$ during the simulation. (C) Evolution of the gravitational potential energy $U_{gra}$, elastic potential energy $U_{ela}$ and kinetic energy $U_{kin}$ during the simulation. }
    \label{fig:2}
\end{figure}

\subsection{Stacking of dipole platelets}

The following problem considers interactions between flat platelets possessing zero charge but nonzero dipole moment. This problem geometry is relevant to multiple physical systems, e.g. liquid crystals, biomineral and colloid platelets ( see, e.g., \cite{Mertelj2013, Smalyukh2010}). We will consider the system of $50$ platelets, consisting of $1094$ pebbles each, placed in a periodic boundary conditions along all the axes. Each platelet is hexagonal, close-packed, two-layer arrangement of spherical pebbles, positively charged layers are highlighted with red, while negatively charged ones are blue. Each layer is therefore consists of $1+3N(N-1)$ pebbles ($N= 14$). The spacing between neighboring pebbles $l = 0.01$, the pebble radius $r=0.008$. The pebble mechanical interaction model is the same as in the previous examples. The platelets are initially  co-oriented, and placed randomly without overlaps in a unit box. 

\begin{figure} 
    \centering
    \includegraphics[width=\textwidth]{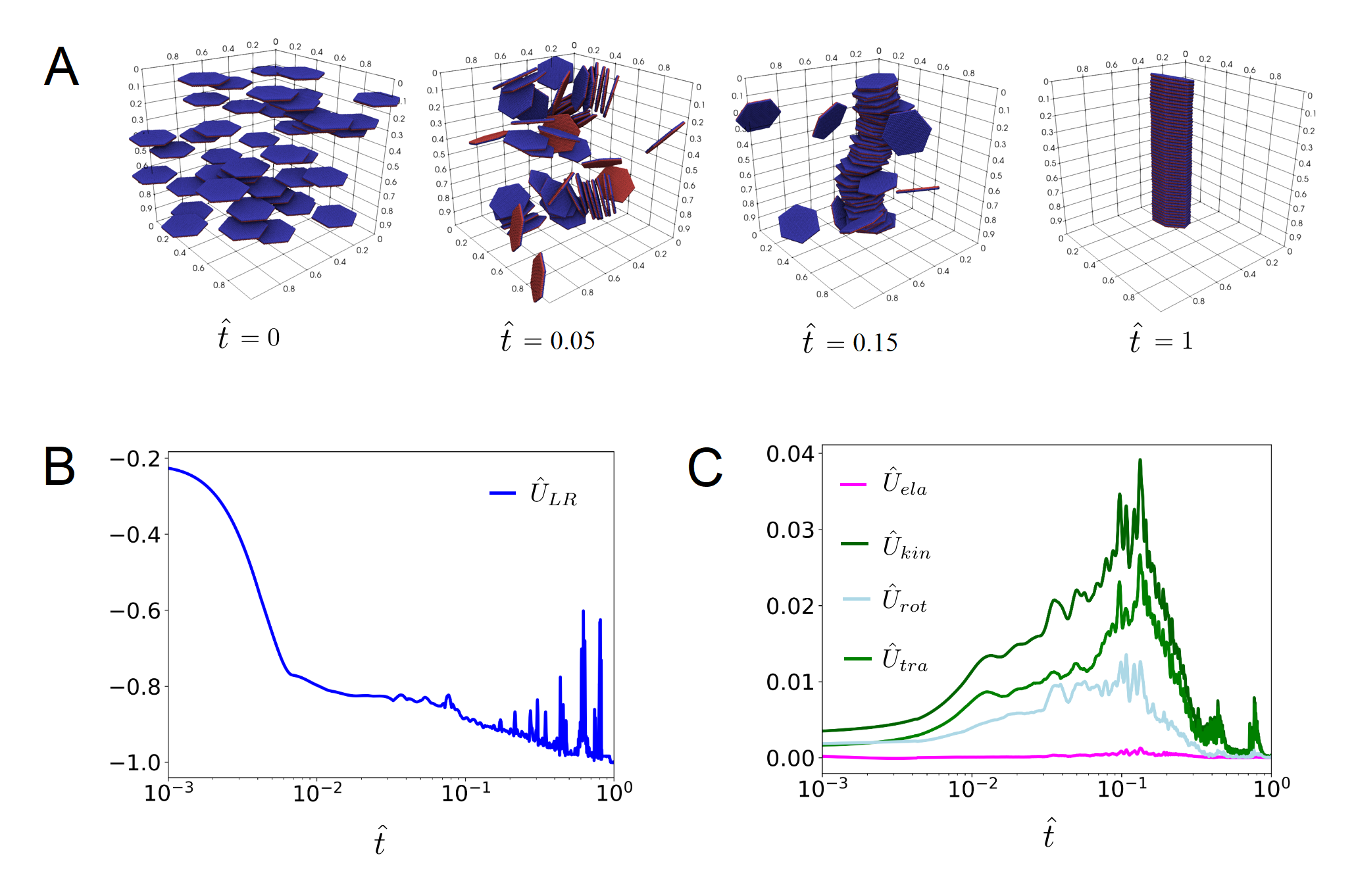}
    \caption{Stacking of dipole platelets. (A) Snapshots of the system at $\hat{t} = 0$, $\hat{t} = 0.05$, $\hat{t} = 0.15$, $\hat{t} = 1$. (B) Time evolution of the potential energy of long-range interactions $U_{LR}$ during the simulation. (C) Evolution of translational ($U_{tra}$), rotational($U_{rot}$) and total ($U_{kin}$) kinetic energy, and the energy of elastic deformations ($U_{ela}$) during the simulation.}
    \label{fig:2}
\end{figure}

As can be seen in Figure 3(A), the simulation features an impressive self-assembly of the dipoles into nearly ideal stack, passing through the periodic boundaries of a simulation cell. It is interesting to observe that the stack is slightly sheared as the equilibrium thickness of all platelets without shear is somewhat less than the distance between periodic boundaries. The observed evolution characterized by highly dynamic, large-scale rearrangements of platelets. Figure 3(B, C) highlights the corresponding evolution of the total energy terms. One notable feature is a substantial fraction of rotational kinetic energy of platelets in a general energy balance. Similarly to the previous example, $U_{ref} = |U_{LR}(1)|$. 

This example can serve as the starting point for any simulation involving a soft matter with unscreened dipole interactions between particles. 
The showcased example took 20 hours on a single-CPU node. The simulation parameters can be found in \texttt{Drivers/LongRange/Clay/}.

\subsection{Gravitational collapse of elastic particles}

Consider $10^5$ identical spherical particles, interacting with each other via the gravitational force 

\begin{equation} \label{electorstatic}
\mathbf{F_{ij}} = \mathbf{r_{ij}}G \frac{m_i m_j}{r_{ij}^3}
\end{equation}     

at large distances and exhibiting dissipative elastic, frictional collisions at short distances. The initial positions of particles are uniformly random in a cube $(0.2, 0.8)^3$. The sides of the cube $(0, 1)^3$ are the elastic walls ensuring that particles do not leave the elementary domain. Note that the periodic boundary conditions can not be used with gravitational potential, as this will lead to divergence of the total potential energy. 

\begin{figure} 
    \centering
    \includegraphics[width=\textwidth]{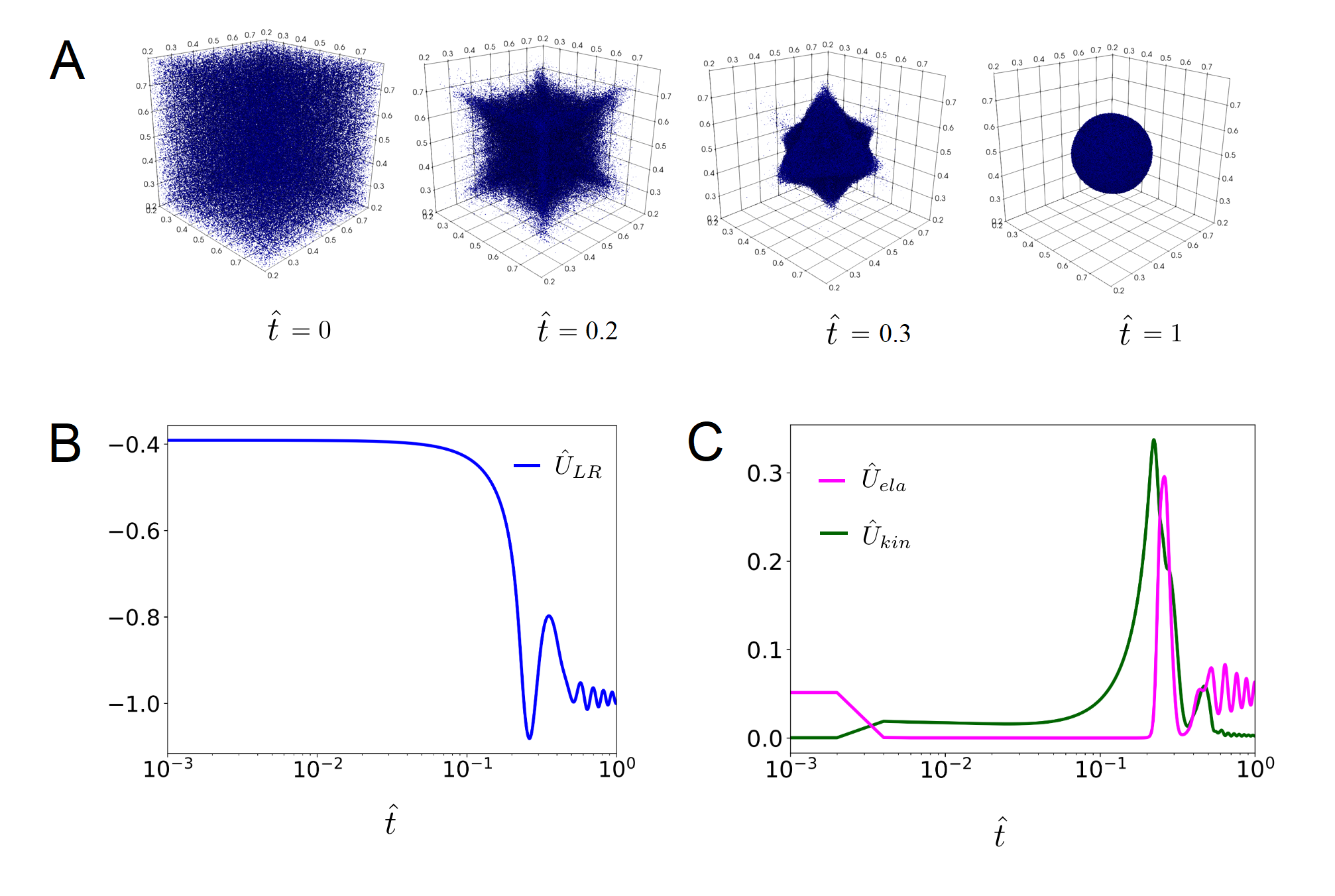}
    \caption{Gravitational collapse of elastic particles. (A) Snapshots of the system at $\hat{t} = 0$, $\hat{t} = 0.2$, $\hat{t} = 0.3$, $\hat{t} = 1$. (B) Time evolution of the potential energy of long-range interactions $U_{LR}$ during the simulation. (C) Evolution of the total kinetic energy ($U_{kin}$), and the energy of elastic deformations ($U_{ela}$) during the simulation.}
    \label{fig:2}
\end{figure}

Figure 1 highlights several snapshots of the evolution of the system. After a several periods of fast decaying oscillations, particles collapse to a nearly spherical aggregate (``planetoid''), whose equilibrium is characterized by the balance between the forces of elastic (i.e., electromagnetic) and gravitational nature. This example may serve as a starting point for a variety of a more complex celestial mechanics simulations - see, e.g., \cite{Vogelsberger2020, Kegerreis2022}.  

This multiphysics simulation with $10^5$ particles, involving $10^5$ timesteps, took only $12$ hours on a regular desktop machine. Similar simulation with $10^6$ particles also remained well within the capabilities of a single desktop machine, demonstrating the performance comparable with the one of the DEM code without KIFMM module.  
The simulation parameters can be found in \texttt{Drivers/LongRange/Galaxy/}.

\section{Conclusion}

This work demonstrated the capabilities of a modeling framework that combines the discrete element method for short-range mechanical interactions with the kernel-independent fast multipole method for long-range interactions. Representing aggregates of charged spherical particles as rigid clumps enables accurate treatment of complex, moment-transferring long-range forces while preserving simple interaction kernels.

The current implementation successfully handles problems involving celestial dynamics, electrostatic interactions, and dipole–dipole coupling. These examples illustrate that the approach can address a diverse set of multiscale particle systems.

The scope of the framework can be extended further. A key direction is the KIFMM-accelerated solution of boundary integral equations. Incorporating BIE methods would enable self-consistent modeling of dielectric polarization of powder particles in external fields (which is the central problem for this class of simulations), magnetostatic analogs of these formulations, and coupled interactions between rigid particles and deformable continua, such as those arising in composite materials (elasticity BIE) or particle-laden flows (Stokes BIE). Combining these capabilities would yield a general multiphysics simulation environment suitable for a broad range of scientific and engineering applications.

In addition, the framework offers substantial opportunities for improving parallel performance and scalability.

All codes described in this work are available at \url{https://bitbucket.org/mercurydpm/mercurydpm/branch/KIFMM_plus_selftests}; the examples described above can be found in \texttt{/Drivers/LongRange/}.

\bibliographystyle{acm}
\bibliography{sample}

\end{document}